# Observation of the Surface Layer of Lithium Metal using *In Situ* Spectroscopy

Ambrose Seo,[1,*] Andrew Meyer,[2] Sujan Shrestha,[1] Ming Wang,[2] Xingcheng Xiao,[3] and Yang-Tse Cheng[1,2]

[1] Department of Physics and Astronomy, University of Kentucky, Lexington, Kentucky 40506, USA
[2] Department of Chemical and Materials Engineering, University of Kentucky, Lexington, Kentucky 40506, USA
[3] General Motors Global Research and Development Center, Warren, Michigan 48090, USA

## Abstract

We have investigated the surface of lithium metal using x-ray photoemission spectroscopy and optical spectroscopic ellipsometry. Even if we prepare the surface of lithium metal rigorously by chemical cleaning and mechanical polishing inside a glovebox, both spectroscopic investigations show the existence of a few tens of nanometer-thick surface layers, consisting of lithium oxides and lithium carbonates. When lithium metal is exposed to room air (~50% moisture), *in situ real-time* monitoring of optical spectra indicates that the surface layer grows at a rate of approximately 24 nm/min, presumably driven by an interface-controlled process. Our results hint that surface-layer-free lithium metals are formidable to achieve by a simple cleaning/polishing method, suggesting that the initial interface between lithium metal electrodes and solid-state electrolytes in fabricated lithium metal batteries can differ from an ideal lithium/electrolyte contact.

* E-mail: a.seo@uky.edu

Recently, metallic lithium (Li) electrodes have attracted attention due to their potential application for lithium metal batteries (LMBs).[1] When used as a negative electrode material for Li-based batteries, the specific capacity of lithium metal can reach 3860 mAh/g, approximately an order of magnitude larger than the theoretical specific capacity limit (i.e., 370 mAh/g) of the graphite electrode used in Li-ion batteries.[2,3] Thus, LMBs are considered a promising candidate for high-energy-density devices, meeting the demands for modern applications including electric vehicles. However, Li metal anode easily forms Li dendrites or mossy structures during electrodeposition, reducing the Coulombic efficiency of a battery cell or short-circuiting it.[1] This problem has hindered the progress of LMBs to a practical battery application. A mechanically robust, electrically insulating but ionically conductive interface layer formed between a metallic lithium electrode and an electrolyte, the so-called solid electrolyte interphase (SEI), has been considered a possible solution for the practical issues relevant to the lithium metal anode. Many research activities aim to create reliable devices by engineering stable SEI layers, preventing fracture or mechanical breakdown during electrodeposition while suppressing dendrite growth.[4-6] Nevertheless, there are some inconsistencies between experimental observations concerning the mechanism of SEI formation. Lithium metal is moisture and air-sensitive, and it will naturally form some passivation layers when exposed to the environment. We note that theoretical modeling of SEI's microstructure and formation assumes an ideal lithium metal surface in general. Hence, understanding the surface state of lithium metal, which is prepared in laboratories, is essential for the progress of LMB research.

Here we report that at least a few nanometer-thick lithium carbonate layers commonly exist on lithium metal surfaces. Even if we perform a thorough surface-preparation of lithium metal, i.e., cleaning and mechanical polishing inside a glove box or a dry room, the nanometer-thick surface layer is observed by x-ray photoemission spectroscopy (XPS) and optical spectroscopic ellipsometry (OSE). Moreover, when lithium metal is exposed to ambient conditions, the surface layer grows at a rate of approximately 24 nm/min according to real-time monitoring of *in situ* optical spectra. Our results indicate that achieving a surface-layer-free lithium metal is a rather formidable task without a chemical etching or physical removal process. Because of this, the initial interfacial states between lithium metal electrodes and solid-state electrolytes in most fabricated solid-state LMBs can be far from a perfect lithium/electrolyte contact. It can also influence the SEI layers formed in the conventional LMBs with liquid electrolytes since the surface state of lithium metal will significantly impact the electrolyte reductions.

We prepared mirror-like surfaces with 0.75-mm-thick lithium metal foils (Alfa Aesar), as shown in Fig. 1(a), by mechanical pressing/polishing process inside an argon-filled glove box (MBraun, oxygen/moisture level < 10 ppm). Since lithium metal is very soft, we polished its surface carefully using conventional laboratory paper (Kimwipes) with dimethyl carbonate solvent, which can remove thick surface layers. During the surface preparation, we used microscope glass plates, which had been cleaned and baked under vacuum at 110 °C in the glovebox anti-chamber, to exert a gentle pressure on a lithium sample to flatten its surface, resulting in a clean mirror-like finish (Fig. 1(a)). Note that this surface preparation process is more rigorous than the method commonly used in laboratories. All lithium samples were transferred into a special argon-filled sample enclosure/holder inside the glovebox to minimize any exposure to air while transferring into external measurement systems such as XPS and OSE.

The chemical composition of the surface layer was revealed using XPS measurements. The $K_\alpha$ x-ray photoemission spectra of (a) Li 1s (b) C 1s, and (c) O 1s orbitals are observed, as shown in Fig. 1(b), indicating that these elements exist even after the rigorous surface preparation in the glovebox. By using Ar-ion milling (with 2000 eV energy and each 20 seconds cycle), which removed the surface layer slowly, the *in situ* XPS data showed noticeable shifts in the binding energies for both the Li 1s (i.e., from 55.2 eV to 53.5 eV) and the O 1s (i.e., from 531.5 eV to 528.2 eV) levels, whereas the peak intensities of the C 1s level decreased monotonously without a peak-shift, as shown in Fig. 1(b). These values of binding energies are consistent with those of lithium carbonate ($Li_2CO_3$) and lithium oxide ($Li_2O$), respectively.[7] (See Fig. S1 in the Supplementary material for determining the peak binding energies.) Moreover, the decrease of the peak intensities of the C 1s level (i.e., 285.0 eV and 289.8 eV) as a function of Ar-ion milling is evidence of diminishing $Li_2CO_3$. Thus, our *in situ* XPS results with Ar-ion milling indicate that the surface layer consists of both $Li_2CO_3$ and $Li_2O$, near the top and the bottom of the surface layer, respectively. However, the XPS data do not give us the thickness information of the surface layer since the precise etching rate of the Ar-ion milling is unknown for this surface layer on lithium metal. Hence, we applied *in situ* OSE to characterize the lithium metal samples.

The OSE results, shown in Fig. 2, are consistent with the XPS results discussed above, suggesting that 30-50 nm thick surface layers commonly exist in a few different lithium metal foils even though their surfaces are prepared inside an Ar filled glovebox, as described previously. Each lithium metal sample was transferred to the vacuum chamber (base pressure of ~$10^{-7}$ Torr) equipped with *in situ* OSE (Woollam M-2000-210 *in situ* option) while minimizing exposure to air. The instrument consists of the light source and analyzer assemblies mounted to the vacuum chamber with an incident angle of 65° relative to the sample's normal axis.[8] Each experimental ellipsometric angle spectrum, $\Psi(\omega)$ and $\Delta(\omega)$, was taken for 2 seconds by averaging 80 dynamic scans every 25 ms in the spectral range of 210 – 1000 nm in wavelength (1.2 – 5.9 eV in photon energy). Ψ and Δ represent the differential changes in amplitude and phase experienced upon reflection by the two *p*- and *s*-polarization components, expressed by the ratio of Fresnel reflection coefficients ($\tilde{R}_p$ and $\tilde{R}_p$): $\rho \equiv \frac{\tilde{R}_p}{\tilde{R}_s} = \tan\Psi\, e^{i\Delta}$. Note that the spectroscopic ellipsometry is a self-normalizing technique, meaning there is no need for reference measurements, and the dielectric function of our samples, $\tilde{\varepsilon}(\omega) = \varepsilon_1(\omega) + i\varepsilon_2(\omega)$, is obtained from Ψ($\omega$) and Δ($\omega$) without a Kramers-Kronig transformation.[9] Since this technique is based on the interference of the *p*- and *s*-polarized lights, the obtained thickness values of the surface layers are very accurate (i.e., sub-Angstrom scale) far better than the diffraction limit of light.

Figure 2(a) shows the ellipsometric angles (Ψ and Δ) of a lithium metal sample. To extract the dielectric function of the surface layer, we used an isotropic layer/semi-infinite-substrate model, as shown in Fig. 2(b), from CompleteEASE (Woollam) software, which derives the solutions of the multiple Jones matrices by a numerical iteration process. For the dielectric function of lithium metal, we used the data from Palik[10] as the initial values for the numerical iteration. Figure 2(c) shows the dielectric functions of the surface layer and the lithium metal underneath. While the dielectric function of lithium metal is consistent with the free-electron Drude model, the dielectric function of the surface layer is quite distinct from that of lithium metal. From the shape of the dielectric function, we can see that the surface layer is quite opaque with low bandgap energies such as lithium carbonate ($Li_2CO_3$) or its composites, which are consistent with the XPS results.

It is also important to investigate how the surface layer would change when the sample is exposed to the ambient condition. Ψ($\omega$) and Δ($\omega$) change drastically as a function of time after lithium metal is exposed to air (humidity ~50%), as shown in Fig. S2. From the raw spectra, it is visible that an interference pattern (i.e., a wavy feature) develops until the exposure time reaches around 5 minutes, implying that the thickness of the surface layer increases gradually. However, after 5-6 minutes of exposure to air, the spectra with short-wavelength light (i.e., photon energies of 5-6 eV) become very noisy due to increased scattering of light from the rough surface. We applied the same isotropic layer/semi-infinite-substrate model to fit the spectra and extracted the thickness (Fig. 3(a)) and the optical conductivity spectra (Fig. 3(b)) of the surface layer as a function of the exposure time. Up until the 5-minute mark, the thickness of the surface layer estimated by *in situ* OSE increased at a rate of approximately 24 nm/min. After 5 min, the layer developed a significantly rough surface, and the increment slowed down. Considering the data up to 5 minutes after which it developed a surface too rough to measure the thickness experimentally, the growth of the surface layer seemed to be governed by *interface-controlled* processes (i.e., *thickness* ∝ *time*) rather than diffusion-controlled processes (i.e., *thickness* ∝ (*time*)$^{1/2}$), as shown in Fig. 3(a).[11] We admit that the current set of data has considerable uncertainty to claim one over the other. However, by assuming a planar interface between lithium metal and the initial surface layer, one can imagine that oxygen or water molecules from the air should move into and across the interface region to increase the thickness of the surface layer. This process requires thermal energies since the molecular motion is short-range. Therefore, we suggest that temperature-dependent measurements will further clarify the mechanism of the formation of the surface passivation layer of lithium metal. It is also worthy of noting that the optical conductivity ($\sigma_1$) spectra of the surface layer, obtained using the relation of $\sigma_1 = \frac{\varepsilon_2 \cdot \omega}{4\pi}$, change systematically with an isosbestic point at around 2.4 eV as the exposure time to air increases (Fig. 3(b)). The increased spectral weight below 2.4 eV is consistent with the commonly seen dark surface color of lithium metal due to our eye's visible range being 1.7 eV - 3.2 eV.

Our *in situ spectroscopic* investigations of both XPS and OSE give indispensable information about the existence of the surface passivation layer of lithium metal even after rigorously preparing its surface inside a glovebox. Hence, the potential impact of this lithium passivation layer should be taken into account when understanding the interface phenomena between lithium metal and solid-state electrolytes, including charge transfer, interfacial reactions, and interfacial mechanical stabilities. Lithium oxide ($Li_2O$) and lithium carbonate ($Li_2CO_3$) are insulating materials, of which layers may suppress electrolyte decomposition. As a result, the additional SEI layers formed on the top surface may not be dense enough to protect the lithium metal electrode effectively. On the other hand, lithium carbonate is an electronic insulator but a $Li^+$ ionic conductor. It was suggested that glassy $Li_2CO_3$ would be desirable for efficient SEI[12] and $LiF/Li_2CO_3$ composite coating on an anode surface could provide an improved passivation function (i.e., a reduced electron leakage) with increased ionic conduction by forming a space charge layer.[13] We suggest further investigations to understand the actual influence of the surface layer on lithium metal batteries. We also suggest that developing *in situ* or *operando* spectroscopic tools such as what was discussed in this paper will lead to a deeper understanding of not only the interface state between lithium metal and an electrolyte but also the formation of SEI under various conditions and environments.

**Supplementary Material**

Detailed x-ray photoemission spectra (Fig. S1) and spectroscopic ellipsometry data as a function of air exposure time (Fig. S2).

**Acknowledgments**

This work was supported by the Vehicle Technologies Office of the U.S. Department of Energy Battery Materials Research (BMR) Program under Contract Number DD-EE0008863. A.S. acknowledges the support of National Science Foundation Grant No. DMR-2104296 for OSE measurements and analyses. A.S. and Y.T.C. acknowledge support from the University of Kentucky Energy Research Priority Area program. A.S. thanks Hian for the careful corrections of this manuscript.

**Figure Captions**

**Fig. 1. (a)** Photographs of one of the lithium metal foils as its surface is cleaned and polished inside an Ar-filled glovebox. **(b)** X-ray photoemission spectra at the K-edge of Li, O, and C, respectively, as its surface layer is slowly removed by the Ar-ion milling process.

**Fig. 2. (a)** Spectroscopic ellipsometry data ($\Psi$ and $\Delta$) taken from a polished lithium metal under vacuum ($10^{-7}$ Torr). The solid lines are the fit curves from the model, whose schematic diagram is shown in **(b).** The thickness of the surface layer is estimated to be about 33 nm with negligible surface roughness. **(c)** The dielectric functions for the surface layer (top) and the lithium metal (bottom).

**Fig. 3. (a)** Thickness of the surface layer as a function of time after the sample is exposed to air. The solid and dashed lines are linear ($\propto t$) and parabolic ($\propto \sqrt{t}$) functions for interface-controlled and diffusion-controlled processes, respectively. **(b)** Optical conductivity spectra of the surface layer as a function of the exposure time. The dashed line is a rough estimate of the direct optical bandgap (1.4 eV) of the initial surface layer.

# References

[1] Xin-Bing Cheng, Rui Zhang, Chen-Zi Zhao, and Qiang Zhang, Chemical reviews **117** (15), 10403 (2017).

[2] Chengcheng Fang, Jinxing Li, Minghao Zhang, Yihui Zhang, Fan Yang, Jungwoo Z Lee, Min-Han Lee, Judith Alvarado, Marshall A Schroeder, and Yangyuchen Yang, Nature **572** (7770), 511 (2019).

[3] B Simon, S Flandrois, K Guerin, A Fevrier-Bouvier, I Teulat, and P Biensan, Journal of power sources **81**, 312 (1999).

[4] Nian-Wu Li, Ya-Xia Yin, Chun-Peng Yang, and Yu-Guo Guo, Advanced materials **28** (9), 1853 (2016).

[5] Wei Liu, Pengcheng Liu, and David Mitlin, Advanced Energy Materials **10** (43), 2002297 (2020).

[6] Mingguang Wu, Yong Li, Xinhua Liu, Shichun Yang, Jianmin Ma, and Shixue Dou, SmartMat **2** (1), 5 (2021).

[7] Kevin N Wood and Glenn Teeter, ACS Applied Energy Materials **1** (9), 4493 (2018).

[8] J. H. Gruenewald, J. Nichols, and S. S. A. Seo, Review of Scientific Instruments **84** (4) (2013).

[9] R. M. A. Azzam and N. M. Bashara, *Ellipsometry and Polarized Light*. (North-Holland, Amsterdam, 1987).

[10] Palik, Handbook of Optical Constants of Solids, Volume II, pp.353-354 (1997).

[11] K. N. Tu, Annual Review of Materials Science **15** (1), 147 (1985).

[12] Mahsa Ebrahiminia, Justin B. Hooper, and Dmitry Bedrov, Crystals **8** (12), 473 (2018).

[13] Jie Pan, Qinglin Zhang, Xingcheng Xiao, Yang-Tse Cheng, and Yue Qi, ACS Applied Materials & Interfaces **8** (8), 5687 (2016).

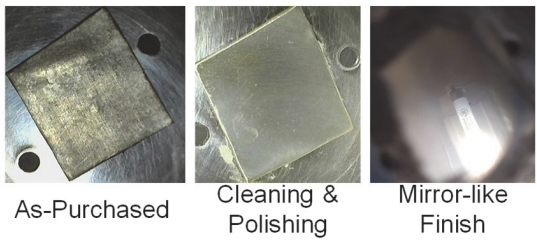
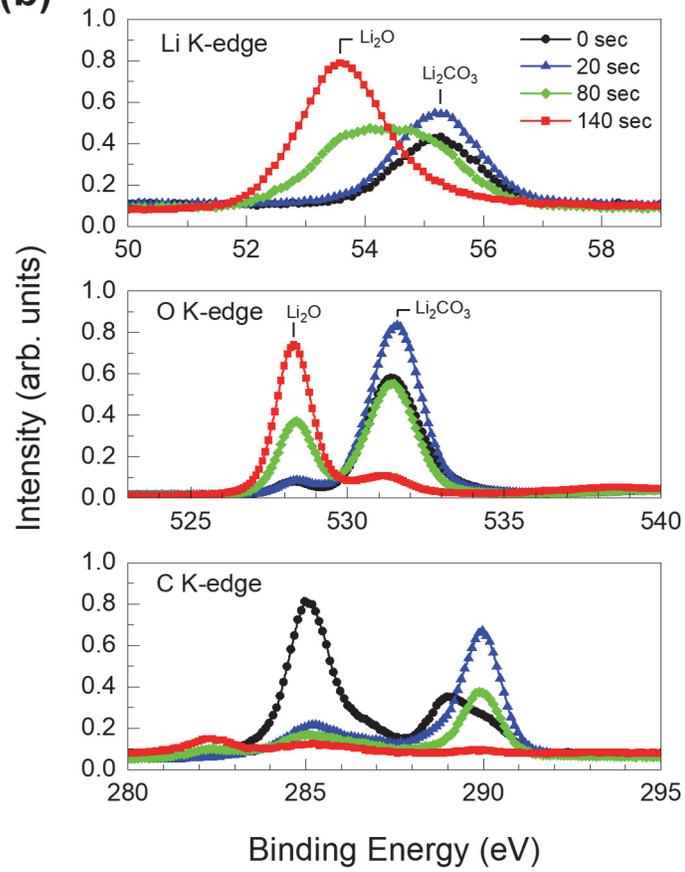

Fig. 1

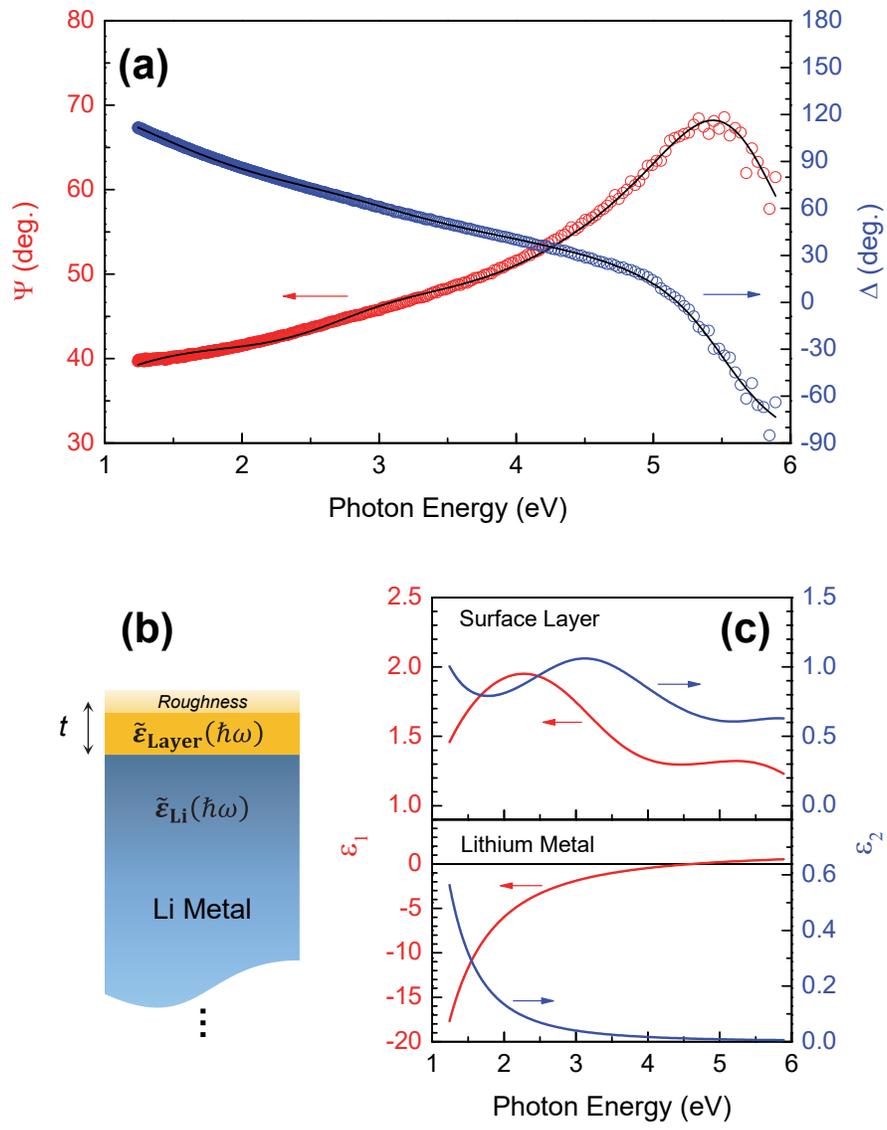

Fig. 2

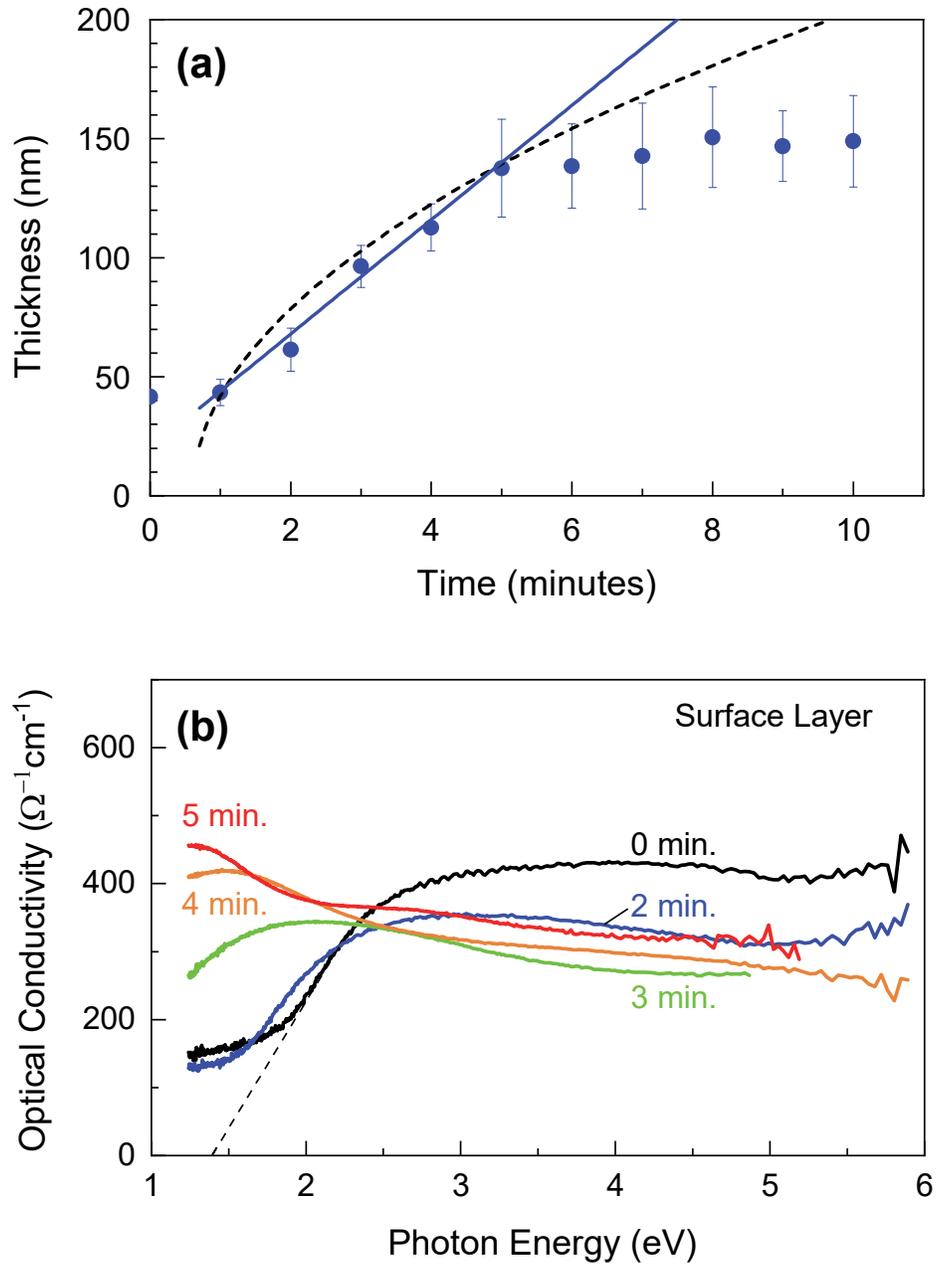

Fig. 3